# Near-field probe of thermal capillary fluctuations of a hemispherical bubble

Z. Zhang[1], Y. Wang[2], Y. Amarouchene[1], R. Boisgard[1], H.Kellay[1], A. Würger[1], A. Maali[1,*]

[1] Univ. Bordeaux, CNRS, LOMA, UMR 5798, F-33405, Talence, France.
[2] School of Mechanical Engineering and Automation, Beihang University, 37 Xueyuan Rd., Haidian District, Beijing, China, 100191

**Abstract:**

We report measurements of resonant thermal capillary oscillations of a hemispherical liquid gas interface obtained using a half bubble deposited on a solid substrate. The thermal motion of the hemispherical interface is investigated using an atomic force microscope cantilever that probes the amplitude of vibrations of this interface versus frequency. The spectrum of such nanoscale thermal oscillations of the bubble surface presents several resonance peaks and reveals that the contact line of the hemispherical bubble is pinned on the substrate. The analysis of these peaks allows to measure the surface viscosity of the bubble interface. Minute amounts of impurities are responsible for altering the rheology of the pure water surface.

* Corresponding author: abdelhamid.maali@u-bordeaux.fr

Molecules that lie at the interface between two fluids are subject to forces that are different from those that are in the bulk. These forces act so as to minimize the surface energy and give rise to the surface tension of interfaces [1,2]: the energy cost to maintain the phase separation of the fluids. These interfaces are host to thermal fluctuations, which are at the origin of the roughness of the interfaces: the fluctuations of the local positions of molecules distort the shape of the interfaces. This phenomenon described using the notion of thermal capillary waves has been the subject of theoretical studies for several decades [3-13]. When such interfaces are confined by imposing a vanishing velocity at the ends of the interface as in the presence of walls, the fluctuation spectrum presents sharp resonance peaks at well-defined frequencies. This situation is similar to the vacuum fluctuations of the electromagnetic field confined in a cavity, where only modes with frequencies that satisfy the boundary conditions can exist inside the cavity.

Experimental studies of thermal capillary waves are mainly performed using techniques such as X-ray reflectivity [14], surface quasi-elastic light scattering (SQELS) [15,16], optical interferometry [17-19] and high speed video imaging [11,20]. Such techniques can also shed light on the viscoelastic properties of surfaces and interfaces when decorated by surfactants.
These additives, even in minute quantities can alter not only the surface tension of surfaces but render these surfaces rheologically non trivial: such surfaces may acquire a surface elasticity and a surface viscosity [21-31]. Techniques to measure the latter remain either difficult to use or have serious drawbacks [30].

In this Letter we report on measurements of the thermal capillary oscillations on the surface of a bubble deposited on a solid substrate immersed in water. The hemispherical shape of the bubble with its contact line pinned on the substrate renders these oscillations resonant with well-defined peaks in their frequency spectrum. The motion of the interface is studied using a cantilever that measures the vibrations amplitude versus frequency. The spectrum of the thermal excitations of the bubble vibrations shows that the contact line is indeed pinned and allows to measure the surface viscosity at the bubble interface with a good precision.

Figure 1a shows the setup used in our experiments. A micro syringe was used to deposit an air bubble on a glass surface, spin-coated with a polystyrene layer (thickness of 100 nm and roughness of 0.2 nm). The bubble was stable for several hours. During the experiment, we bring an atomic force microscope (AFM) cantilever in contact with the air bubble surface and measure its time-dependent position, from which the spectral density of the capillary fluctuations of the water-air interface was determined. The radius $R$ and contact angle of the bubble were determined from the top-view (Fig.1b) and side view (Fig.1c) optical images, respectively.

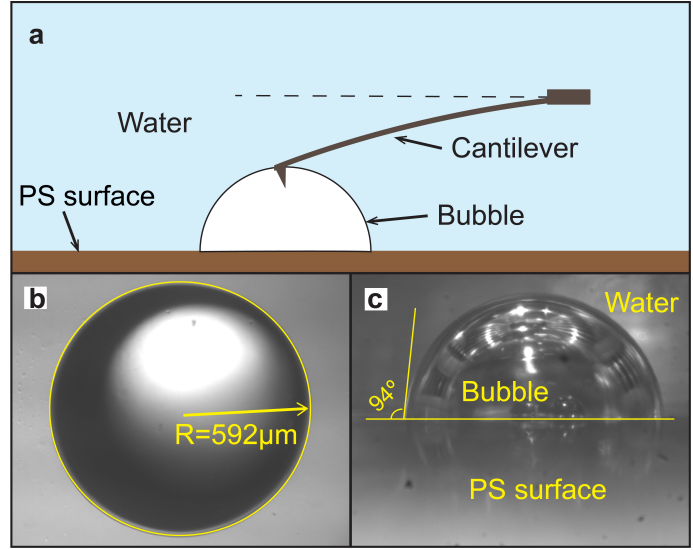

Fig.1 **a)** *Experimental setup. The bubble was deposited on polystyrene (PS) surface, and the cantilever tip was used to probe the vibration of the bubble.* **b)** *Top view and* **c)** *side view images, from which we obtain the bubble radius $R = 592 \pm 5$ μm and the contact angle of $94 \pm 2°$.*

Experiments were performed using an AFM (Dimension 3100, Bruker) equipped with a liquid cell, (DMFT-DD-HD) that allows operation in a liquid, and two different cantilevers (MLCT, Bruker, stiffness $k_c = 0.024 \pm 0.002$ N/m ) and (CSG01, NT-MDT, stiffness $k_c = 0.12 \pm 0.02$ N/m ). The position of the cantilever was controlled by the AFM stepping motor stage allowing to bring the tip in contact with the north pole of the bubble. Once this contact was established, the cantilever was driven solely by the vibrations of the bubble. The maximum amplitude of these vibrations was $< 1\ nm$. The vertical deflection of the cantilever, due to these oscillations, was acquired by an A/D acquisition board (PCI-4462, National Instrument, USA). From this time series of cantilever deflection signal, the power spectral density (PSD) of the AFM cantilever vibrations was calculated.

Typical PSD curves are shown in Fig. 2a. The blue curve was measured for the cantilever in bulk water, far from the bubble, and shows clearly the vibrational mode of the cantilever with characteristic frequency near $4$ kHz. The main driving force for these cantilever fluctuations is the thermal noise [32-35]. The red curve depicts the PSD of the cantilever displacements when in contact with the bubble.

As a main finding, we observe well-defined resonance peaks for the cantilever in contact with the bubble. While the exact origin of these fluctuations is a priori not known, we postulate that the displacement fluctuations of the cantilever are driven by the thermal fluctuations of the bubble surface.

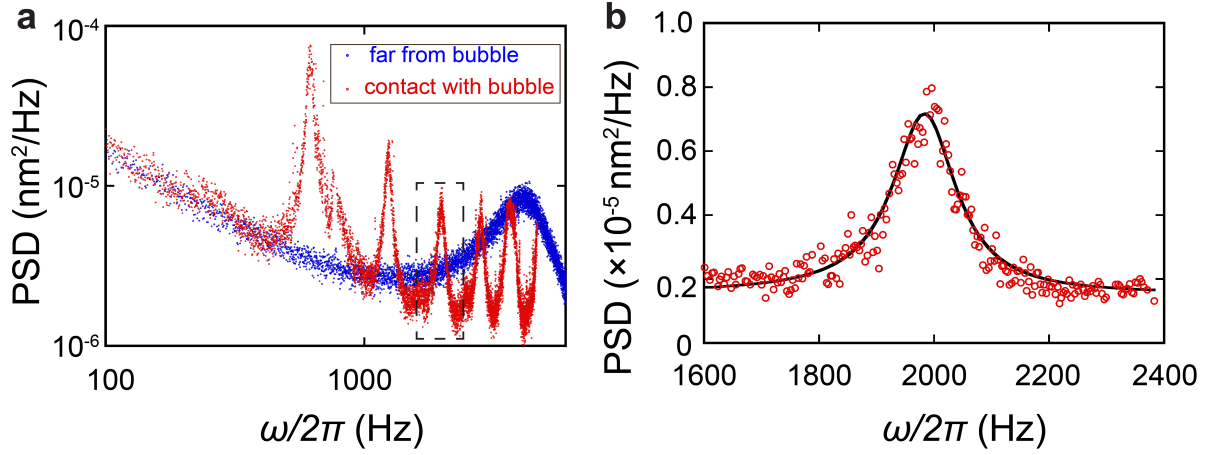

Fig. 2: Example of the measured PSD Curves using a cantilever with stiffness $k_c = 0.12$ N/m. *a) The thermal spectra of the cantilever far from the bubble (blue circles) and in contact with the bubble (red circles) deposited on PS surface. b) The spectrum (circles) and the fitting curve using Eq. 2 (solid line) for the third peak in a).*

For a spherical inviscid liquid drop of surface tension $\sigma$, radius $R$, and fluid density $\rho$, Rayleigh predicted that the square of the resonance frequencies of the drop oscillations are multiples of [36]:

$$\omega_0^2 = \frac{\sigma}{\rho R^3}. \tag{1}$$

For such drops or bubbles, each vibrational mode can be described as an oscillating string with an amplitude $\xi_n(t)$, satisfying the equation of motion

$$m_n(\ddot{\xi}_n + 2\beta_n \dot{\xi}_n + \omega_n^2 \xi_n) = \mathcal{F}_n(t), \tag{2}$$

with the effective mass $m_n$, the damping coefficient $\beta_n$, the resonance frequency $\omega_n$ which is some multiple of $\omega_0$ and the mode number $n$. We use the shorthand notation $\dot{\xi}_n = d\xi_n/dt$. As we postulated above, the driving force $\mathcal{F}_n(t)$ is due to thermal noise, which is assumed uncorrelated in time and independent for each mode. Taking the Fourier transform of Eq. (2) and using $|\mathcal{F}_n(\omega)|^2 = 2\beta_n m_n k_B T$, we obtain the one sided power spectral density $\text{PSD} = \sum_n |\xi_n(\omega)|^2$ in the form

$$\text{PSD}(\omega) = \sum_{n \geq 1} \frac{4\beta_n}{(\omega^2 - \omega_n^2)^2 + 4\beta_n^2 \omega^2} \frac{k_B T}{\pi m_n}. \tag{3}$$

Since the measured contact angle of our bubbles is very close to 90°, they may be considered hemispherical. The axisymmetric modes of a free bubble are described by Legendre polynomials of even degree, $P_{2n}(\cos \theta)$ with the polar angle $\theta$ [36], which are labelled as $n = 1, 2, 3, ...$

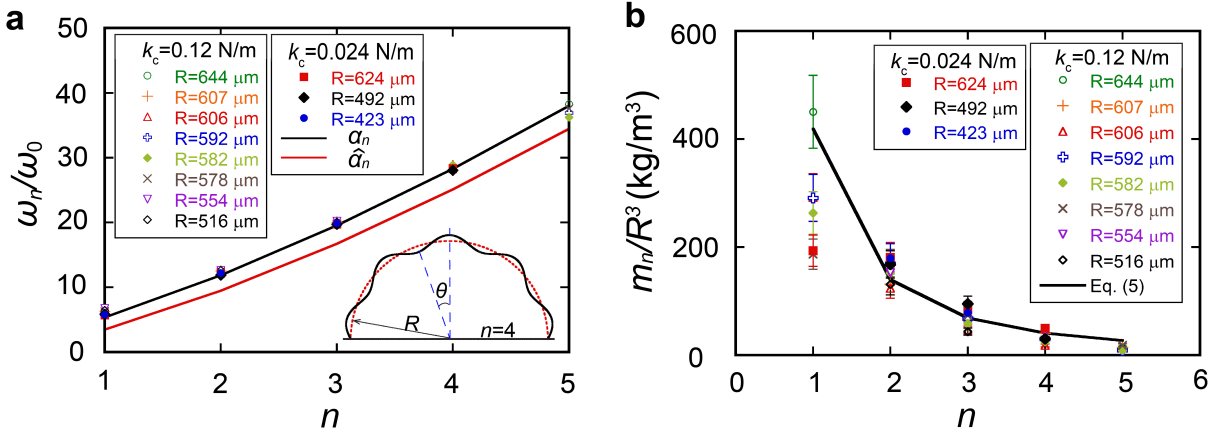

Fig. 3 a) *The results of the normalized resonance frequencies $\omega_n/\omega_0$ of the bubble versus the mode numbers. The black solid line connects the resonance frequency for pinned contact lines, and the red solid line for a freely moving contact line, as given in the main text. b) The results of the effective mass normalized by the cubic power of the radius of the bubble versus the mode number for different bubbles. The dots with different colors and shapes represent the different measurements for different bubbles. The black line represents the theoretical results which was given by Eq. (5).*

The solutions of odd degree are not compatible with the presence of the solid boundary, since they would imply a finite velocity normal to the solid surface.

The vibrational spectrum depends on the boundary conditions for the oscillation amplitude at the solid surface. If the contact line moves freely along the surface, the capillary oscillations $\xi(t, \theta) = \sum_n \xi_n(t) P_{2n}(\cos\theta)$ are simply the even modes of a spherical bubble in an inviscid liquid, where the natural frequencies $\hat{\omega}_n = \hat{\alpha}_n \omega_0$ are given by $\hat{\alpha}_n = \sqrt{(2n-1)(2n+1)(2n+2)}$ [36-38]. In Eq. (1) it is the density of the fluid surrounding the bubbles that is used [36,37,39,40].

For a pinned contact line, on the other hand, the non-slip boundary condition imposes the global constraint $\xi(t, \theta = \frac{\pi}{2}) = 0$. Then the vibration amplitude of a single mode is a superposition of even Legendre polynomials, $\xi_n(t) = \sum_k b_{nk} P_{2k}(\cos\theta)$, with $b_{nk} < b_{nn}$. For contact angle 90° Lyubimov et al. calculated the resonance frequencies $\omega_n = \alpha_n \omega_0$, where the coefficients $\alpha_n$ are solutions of the implicit equation [39,41,42]

$$\sum_{k \geq 1} \frac{(2k+1)(4k+1)}{\hat{\alpha}_k^2 - \alpha_n^2} P_{2k}(0)^2 = 0. \qquad (4)$$

The above description of bubble oscillations extended to the case of a half bubble resting on a solid surface with contact angle close to 90° can be compared quantitatively to the results of Fig. 2a.

Figure 2b shows a fit to one of the peaks of the PSD using Eq. (3). Such fits, which account quantitatively for the shape of the peaks, allow to determine the resonance frequency as well as the effective mass and the damping coefficient for the different mode numbers *n*. The values for the resonance frequencies normalized by $\omega_0$ obtained from the PSDs of different bubbles are plotted in Fig. 3a. The results are best accounted for using nonslip boundary conditions; contact line pinning stiffens the vibrations and enhances the frequencies with respect to those obtained for slip boundary conditions, $\alpha_n > \hat{\alpha}_n$. Indeed, the comparison of Fig. 3a where the theoretical values for both non-slip and slip boundary conditions are displayed along with the experimental values, leads to the conclusion that the contact line of our bubble does not move on the surface but is pinned on the substrate. In the following we always refer to $\omega_n = \alpha_n \omega_0$.

Figure 3b shows the effective mass $m_n$ versus mode number $n$. These masses are extracted from fits of the PSD to Eq. (3). Here data from different bubble radii are displayed.

Following Rayleigh [36], we express the kinetic energy through the velocity potential and integrate over the fluid volume [37,39,43] to obtain the effective mass:

$$m_n = \frac{2\pi \rho R^3}{(2n+1)(4n+1)}. \tag{5}$$

A comparison of Eq. (5) with the data for the measured effective mass $m_n$ is shown in Fig. 3b. The effective mass has been normalized by $R^3$ giving a reasonable collapse of the data from different experiments using different bubble radii. Apart from the mode at $n = 1$, the data from different realizations collapse on a single curve. Further, the decrease of this effective mass with the mode number $n$ as anticipated by Eq. (5) is accounted for quantitatively. While our assumption of independent modes seems plausible for higher modes, it is not for the first mode mainly because its frequency is intermediate between the first and second free modes contrary to the higher modes whose frequencies are rather close to the corresponding free mode (see table in SI).

So far, the bubble vibrations have been treated in the framework of potential flow of an inviscid fluid. Now we turn to the damping coefficient $\beta_n$, which is directly related to the width of the peaks in Fig. 2a. Damping of capillary waves has been extensively studied, and various mechanisms have been identified. At clean interfaces, viscous damping is the dominant source of dissipation [40], arising essentially from the flow in a surface layer of thickness $\delta = \sqrt{\eta/\rho\omega}$, where $\eta$ is the viscosity of the fluid surrounding the bubble. For frequencies in the kHz range, the penetration depth $\delta$ is of the order of ten microns, much smaller than the radius of our bubbles. In such a case,

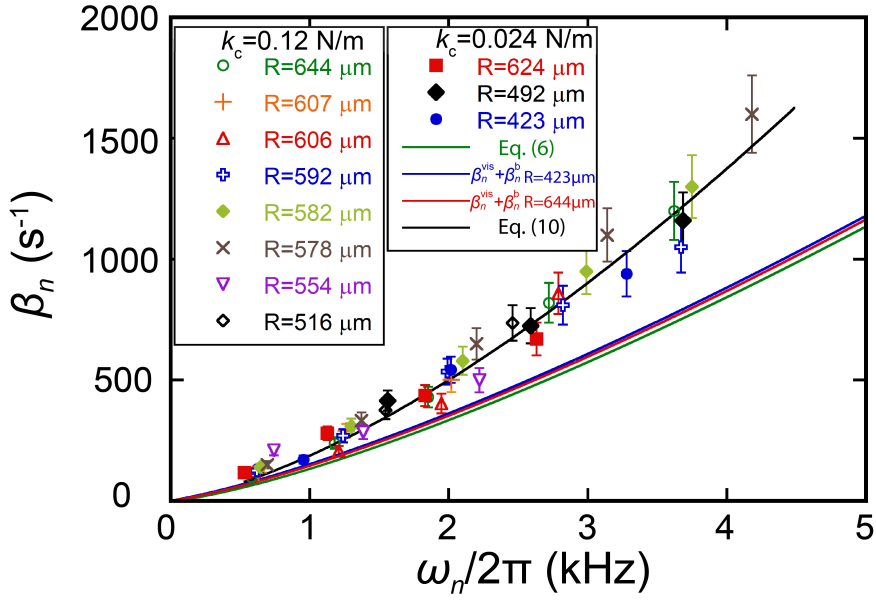

*Fig. 4: Damping versus the frequency for different bubbles. The green solid line corresponds to the viscous damping $\beta_n^{\mathrm{vis}}$ as in Eq. (6). The red and blue solid lines correspond to the viscous damping plus boundary damping ($\beta_n^{\mathrm{vis}} + \beta_n^{\mathrm{b}}$), where $\beta_n^{\mathrm{b}}$ were calculated for $R = 644$ μm and $= 423$ μm respectively by Eq. (7). The black solid line is calculated from Eq. (10) and accounts for all terms of viscous damping $\beta_n^{\mathrm{vis}}$ (Eq. (6)), boundary damping $\beta_n^{\mathrm{b}}$ (Eq. (7)) and the effects of surfactants $\beta_n^{\mathrm{s}}$ (Eq. (8)), with the surface viscosity is $\eta_s = (1.5 \pm 0.2) \times 10^{-7}\ Pa \cdot s \cdot m$.*

Rayleigh's dissipation function is readily evaluated, resulting in the viscous damping coefficient $\beta_n^{\mathrm{vis}}$ [37,39,44,45]:

$$\beta_n^{\mathrm{vis}} = \frac{2\eta}{\rho^{1/3}\sigma^{2/3}}\omega_n^{4/3}. \tag{6}$$

The above expression for the damping coefficient does not account for dissipation on the solid substrate [23] nor for the presence of additional surface damping [46]. In a study of Faraday waves [47], Milner assumed non-slip boundary conditions at the container walls [23]; the increase of velocity within the penetration length $\delta$ results in a boundary contribution to the damping rate, $\omega\delta/L$, which is inversely proportional to the container size $L$. This leads us to estimate the boundary damping coefficient $\beta_n^{\mathrm{b}}$ of the bubble oscillations as [39]:

$$\beta_n^{\mathrm{b}} = \frac{3\sqrt{2}\eta^{1/2}}{40\rho^{1/2}R}\omega_n^{1/2}. \tag{7}$$

Besides this additional damping localized on the solid substrate, surface induced damping (either from an intrinsic surface viscosity [46] or from the presence of surface active agents resulting from possible contamination of the surface [21,24,30,31]) provides another source of dissipation. The case of a flat infinite interface, covered partially by contaminants, was studied by Miles, including surface viscosity and elasticity [21,31]. With the parameters of our system, the contribution of surface elasticity to the damping is very small [39], and thus will be neglected. Moreover, if we consider the case of insoluble contaminants only, then the damping coefficient $\beta_n^s$ for capillary waves reads

$$\beta_n^s = \frac{\sqrt{2}\eta^{1/2}\omega_n^{7/6}}{4\rho^{1/6}\sigma^{1/3}} \frac{\varsigma(\varsigma+2)}{2+\varsigma(\varsigma+2)}, \qquad (8)$$

with the dimensionless quantity $\varsigma$ expressed as a function of the frequency:

$$\varsigma = \frac{\sqrt{2}\rho^{1/6}}{\eta^{1/2}\sigma^{2/3}}\omega_n^{5/6}\eta_s. \qquad (9)$$

The surface viscosity $\eta_s$ includes both dilatation and shear viscosities, which cannot be distinguished in our experiment.

Figure 4 shows data obtained for different bubble radii. The damping coefficient is plotted versus the resonance frequency, given by the mode number $n$ and the radius of the bubble $R$. The damping coefficient predicted by Eq. (6) is plotted as a green line in Fig. 4. Although it captures the overall trend of increasing $\beta_n$ versus $\omega_n$, it is about a factor 2 smaller than experimental values. Thus, viscous damping is not sufficient to explain the measurements. Further, additional damping due to the presence of the solid boundary does not contribute significantly. The blue and red lines take into account both viscous damping and boundary damping and are calculated using two different radii spanning the range of explored values in our experiments. The experimental values remain higher than expected from viscous and boundary damping indicating that additional damping is needed. The black solid line in Fig. 4 depicts the total damping coefficient $\beta_n^{\text{tot}}$, by accounting for both bulk and surface viscosities,

$$\beta_n^{\text{tot}} = \beta_n^{\text{vis}} + \beta_n^{\text{b}} + \beta_n^{\text{s}}. \qquad (10)$$

In the above equation, the surface viscosity is the only adjustable parameter, taken as $\eta_s = (1.5 \pm 0.2) \times 10^{-7}$ Pa·s·m. Each of the three contributions to Eq. (10) is necessary for a satisfactory fit of the data from different experiments but the contribution of the surface viscosity is crucial for a better agreement with experimental values.

In order to verify that the flow created by the vibrating cantilever does not alter the dissipation and add an artefact in the damping coefficients, we have performed measurements with two cantilevers of different dimensions, one of stiffness $k_c = 0.024$ N/m, width $w = 22$ μm, length $l = 208$ μm, and a second one with $k_c = 0.12$ N/m, $w = 34$ μm, $l = 350$ μm. Because of the larger dimensions of the second cantilever, any effect of the cantilever beam on the damping coefficient should be significantly stronger than that of the first one. Yet Fig. 4 shows that there is no difference in the damping coefficients, which leads us to the conclusion that there is no unwanted backreaction of the vibrating cantilever on the bubble dynamics.

The fitted value of $\eta_s$ is about ten times larger than those reported by Earnshaw [46] for a pure water surface and Zell et al. [30] for soluble surfactant covered interfaces. The discrepancy could be due to the fact that in the present work, the surface viscosity measured accounts both for surface dilatational viscosity as well as surface shear viscosity. Further, we believe that our measurements are not devoid of surface contamination. In fact, in one of our previous studies [48] despite the fact that a careful protocol was applied to minimize surface impurities, the air water surface was found to be prone to contamination rather quickly with drastic effects on the properties of the water air interface even for minute quantities of contaminants. We believe that there are similar effects here. Remarkably, our experimental technique is capable of probing the surface viscosity with a high precision. This is shown by Fig. 4 where the bulk effects are well below the measured damping rates. We hypothesize that coupling such a technique with precise techniques for measuring surface shear viscosities (such as that of Zell et. al. [30]) provides a reliable technique to pin down the surface rheology of interfaces with various surface active agents and disentangle dilatational from shear viscosities.

In conclusion, our experiments demonstrate that the AFM cantilever technique developed here is a powerful tool to probe the thermal capillary fluctuations of bubble surfaces. The spectrum of the fluctuations presents sharp resonance peaks for specific frequencies for which the motion of the interface is much more important than for other frequencies. Our measurements demonstrate that the contact line of a half bubble resting on a solid surface is pinned on the substrate and allows us to measure the additional damping due to the presence of minute amounts of contaminants. Moreover, our experimental method provides a useful new tool to probe the surface rheology.